# Effect of three-spin interaction on thermal entanglement in Heisenberg XXZ model

Jing-Heng Fu(傅靖恒), Guo-Feng Zhang(张国锋)[*]

*School of Physics and Nuclear Energy Engineering, Beihang University, Xueyuan Road No. 37, Beijing 100191, China*

**Abstract:** The effect of three-spin interaction $k$ on thermal entanglement between alternate qubits is studied using pairwise concurrence $C$ and energy level diagram. It is found that $k$ breaks the symmetry about the effect of magnetic field $h$ on $C$. It shifts a dip structure and gradually effaces a boot structure when $|k| < |J|$ ($J$ is spin exchange coupling). A peak with $C$ maintains 1 appears and expands, elbowing the dip backwards when $|k| > |J|$. A sudden change of the concurrence occurs around $|k| = |J|, h = -k$. Similar conclusions about nearest-neighbor qubits are directly given.

**Keywords:** Heisenberg XXZ model; concurrence; three-spin interaction

**PACS number(s)**: 03.67.Hk, 03.65.Ud, 75.10.Jm

## I. Introduction

Entanglement is thought to play an important role in quantum information [1, 2]. Thermal entanglement, which refers to the entanglement exists in thermal equilibrium state, has attracted a lot of attention due to its special characteristics. It is usually studied by means of the Heisenberg model, which is a simple but effective description of magnetic systems. Entanglement in spin chains with diverse numbers [3-6], different spin multiplicity [7], as well as the effect caused by anisotropy or other various interactions [8-11] has been investigated.

In the previous studies of thermal entanglement in spin chain system, only the nearest-neighbor spin interaction is considered. Besides the two-spin kind of interactions, however, multi-spin interactions may also exist in spin chains. It has been proved that multi-spin interaction can induce quantum phase transition phenomenon and change the critical point of a spin system [12-14]. Consequently, some physical quantities, such as the entanglement, the quantum discord and the geometric phase, exhibit singular properties at the phase transition point. According to these results, we naturally raise some questions: what does the thermal entanglement behave when there exists multi-spin interaction in the spin chain? In this paper, we will focus on the effect of three-spin interaction on the thermal entanglement of Heisenberg XXZ model.

---

[*] Correspondence and requests for materials should be addressed to G.Z.(gf1978zhang@buaa.edu.cn)

## II. Heisenberg model of a three spin chain

When an external magnetic field $h$ along z axis is homogeneously applied to a three-spin-1/2 chain system, the Hamiltonian considering both the nearest-neighbor-spin coupling strength $J$ and three-spin interaction $k$ reads

$$H_S = J\sum_{i=1}^{2}(\sigma_i^x \sigma_{i+1}^x + \sigma_i^y \sigma_{i+1}^y) + h\sum_{i=1}^{3}\sigma_i^z + k(\sigma_1^x \sigma_2^z \sigma_3^x + \sigma_1^y \sigma_2^z \sigma_3^y). \tag{1}$$

A positive $J$ corresponds to antiferromagnetism while a negative one indicates ferromagnetism.

The eigenstates $\varphi_i$ and eigenvalues $\varepsilon_i$ of $H_S$ along with pairwise concurrence $C_i^{(mn)}$, i.e. the concurrence between the $m$th and $n$th qubits that equals $\text{Tr}_l(|\varphi_i\rangle\langle\varphi_i|)$ in which $l$ is the remaining qubit, are listed below.

$$|\varphi_1\rangle = |000\rangle, \varepsilon_1 = -3h, C_1^{(13)} = 0; \tag{2a}$$

$$|\varphi_2\rangle = |111\rangle, \varepsilon_2 = 3h, C_2^{(13)} = 0; \tag{2b}$$

$$|\varphi_3\rangle = \frac{1}{\sqrt{2}}(-|110\rangle + |011\rangle), \varepsilon_3 = h - 2k, C_3^{(13)} = 1; \tag{2c}$$

$$|\varphi_4\rangle = \frac{1}{\sqrt{2}}(-|100\rangle + |001\rangle), \varepsilon_4 = -h + 2k, C_4^{(13)} = 1; \tag{2d}$$

$$|\varphi_5\rangle = \frac{1}{\sqrt{2}}\sin p\, |100\rangle + \frac{|J|}{J}\cos p\, |010\rangle + \frac{1}{\sqrt{2}}\sin p\, |001\rangle, \varepsilon_5 = -h - k - a, C_5^{(13)} = \frac{4J^2}{a(a-k)}; \tag{2e}$$

$$|\varphi_6\rangle = \frac{1}{\sqrt{2}}\sin q\, |110\rangle - \frac{|J|}{J}\cos q\, |101\rangle + \frac{1}{\sqrt{2}}\sin q\, |011\rangle, \varepsilon_6 = h + k - a, C_6^{(13)} = \frac{4J^2}{a(a+k)}; \tag{2f}$$

$$|\varphi_7\rangle = \frac{1}{\sqrt{2}}\sin q\, |100\rangle + \frac{|J|}{J}\cos q\, |010\rangle + \frac{1}{\sqrt{2}}\sin q\, |001\rangle, \varepsilon_7 = -h - k + a, C_7^{(13)} = \frac{4J^2}{a(a+k)}; \tag{2g}$$

$$|\varphi_8\rangle = \frac{1}{\sqrt{2}}\sin p\, |110\rangle - \frac{|J|}{J}\cos p\, |101\rangle + \frac{1}{\sqrt{2}}\sin p\, |011\rangle, \varepsilon_8 = h + k + a, C_8^{(13)} = \frac{4J^2}{a(a-k)}; \tag{2h}$$

$$C_1^{(12)} = C_2^{(12)} = C_3^{(12)} = C_4^{(12)} = 0, C_5^{(12)} = C_6^{(12)} = C_7^{(12)} = C_8^{(12)} = \frac{2|J|}{a}, C_i^{(23)} = C_i^{(12)}; \tag{2i}$$

where $a = \sqrt{8J^2 + k^2}$, $\sin p = \frac{2\sqrt{2}|J|}{\sqrt{8J^2+(k-a)^2}}$, $\cos p = \frac{k-a}{\sqrt{8J^2+(k-a)^2}}$, $\sin q = \frac{2\sqrt{2}|J|}{\sqrt{8J^2+(k+a)^2}}$ and $\cos q = \frac{k+a}{\sqrt{8J^2+(k+a)^2}}$.

In this paper, the deduction bases on $\varepsilon_i$ and $C_i^{(mn)}$, which are not affected by the sign of $J$. We set $J$ to 1 hereunder, use $h$ and $k$ to delegate ratios $\frac{h}{J}$ and $\frac{k}{J}$ respectively. Concurrences of the nearest-neighbor qubits $C_i^{(12)}$ and $C_i^{(23)}$ have only two different expressions as shown in (2i), causing simpler conclusions comparing to alternate qubits 1 and 3, thus we study the latter in detail and give conclusions of the former directly near the end. For convenience, $C_i$ denotes $C_i^{(13)}$ in the main body of the paper.

From the concurrences we can find that the alternate qubits of eigenstates $\varphi_1$ and $\varphi_2$ keep unentangled regardless of $J, h, k$ while $\varphi_3$ and $\varphi_4$ are just opposite. It is obvious that $C_5 = C_8$ and $C_6 = C_7$. Besides, they are only affected by three-spin interaction $k$ ($J = 1$). The two different concurrences as a function of $k$ is plotted in Fig. 1. When $k$ changes from $-\infty$ to $\infty$, $C_5$ and $C_8$ increase monotonically from 0 to 1, $C_6$ and $C_7$ decrease from 1 to 0. They are all 0.5 when $k$ equals 0.

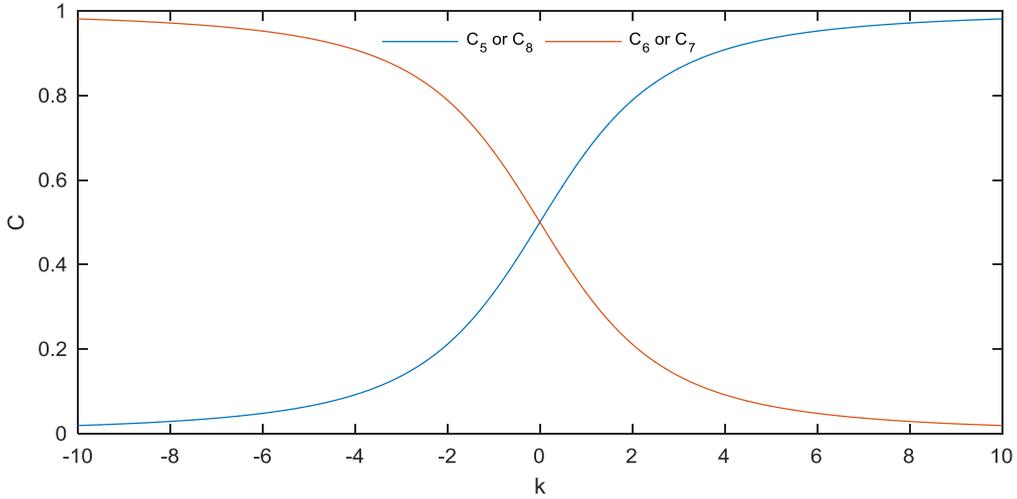

Fig. 1: (Color online) Concurrences of eigenstates $\varphi_5$ to $\varphi_8$ as functions of three-spin interaction $k$. They all have upper bound 1 and lower bound 0. $C_5, C_8$ are monotone increase. $C_6, C_7$ are monotone decrease.

### III. Effect of three-spin interaction $k$ on alternate qubits

We use density operator

$$\rho(T) = \frac{1}{Z}\exp\left(-\frac{H}{k_B T}\right) \tag{3}$$

to describe the state of the system at thermal equilibrium. Partition function $Z = \text{Tr}\left[\exp\left(-\frac{H}{k_B T}\right)\right]$. $k_B$ is Boltzmann's constant, which we will set to 1 hereafter [2]. The reduced density matrix $\rho_{13} = \text{Tr}_2[\rho(T)]$ of the alternate qubit pair 13 has the form

$$\rho_{13} = \frac{1}{Z}\begin{pmatrix} u & 0 & 0 & 0 \\ 0 & w & y & 0 \\ 0 & y & w & 0 \\ 0 & 0 & 0 & v \end{pmatrix}. \tag{4}$$

Let $\beta = \frac{1}{T}$, then $u, v, y, Z$ are

$$u = e^{3h\beta} + \frac{1}{2a}[(a+k)e^{(h+k-a)\beta} + (a-k)e^{(h+k+a)\beta}], \tag{4a}$$

$$v = e^{-3h\beta} + \frac{1}{2a}[(a+k)e^{-(h+k-a)\beta} + (a-k)e^{-(h+k+a)\beta}], \tag{4b}$$

$$y = \frac{1}{2a}\{(a+k)\cosh[(h+k+a)\beta] + (a-k)\cosh[(h+k-a)\beta]\} - \cosh[(h-2k)\beta], \tag{4c}$$

$$Z = u + v + 2y + 4\cosh[(h-2k)\beta]. \tag{4d}$$

According to [5], we can calculate the concurrence of the alternate qubits with equation (superscript is omitted)

$$C = \frac{2}{Z}\max(|y| - \sqrt{uv}, 0). \tag{5}$$

With non-negative three-spin interaction $k$ as parameter, magnetic field $h$ and temperature $T$ as dependent variables, $C$ is plotted in Fig. 2 (a), (b), (c), and (d). When $k = 0$, as shown in Fig. 2(a), there exists two boot-like structures laid out symmetrically on both sides of line $h = 0$. The height of the boots at positive and negative $h$ areas, which is denoted by $C_+$ and $C_-$ respectively and is defined as the maximum concurrence within the boot-like area as $T$ approaches 0, both equals 0.5. The two boots stand very close but do not joint, thus form a dip at the ultralow temperature region. The value of the dip $C_{dip} = \lim_{T \to 0} C|_{h=0} = 0$ in this case. When $k$ increases from 0 to 1, the width, height and length of the positive boot expand, making it occupy part of the negative $h$ area (For brevity, we still call it positive boot with height $C_+$). On the other hand, the negative boot shrinks in all three dimensions (Fig. 2(b)). In the end it disappears at $k = 1$, as shown in Fig. 2(c). The expansion and shrinkage of the two boots result in the dip between them shifting towards negative $h$ direction. When $k$ is larger than 1, a peak whose height $C_-$ maintains 1 suddenly appears at the position the negative boot has just disappeared, then it expands in both width and length, turns to a boot-like structure, and elbows the positive boot back towards the positive area. Consequently the dip shifts back towards positive $h$ direction. Fig. 2(d) exhibits the situation of $k = 1.5$. We can find the symmetry of concurrence about $h$ is broken when $k$ is considered.

This phenomenon can be explained using the energy level diagram of the eigenstates plotted in Fig. 2(e), (f), (g), and (h). At ultralow temperature $T$, the system mainly stays in the ground state. For any given $k$, the ground state energy as a function of $h$ forms a broken line with several inflexion points in the diagram. Point on a segment or half-line of the broken line indicates that the system stays in the eigenstate holding eigenvalue represented by that segment or radial. While the inflexion point, which arises due to crossing of energy levels, corresponds to a mixed state of the degenerate states. Here the alternate qubits' state can be described by the reduced density matrix

$$\rho_{13}^{(i_1 i_2 \cdots i_n)} = \frac{1}{n} tr_3 \big[ |\varphi_{i_1}\rangle\langle\varphi_{i_1}| + \cdots + |\varphi_{i_n}\rangle\langle\varphi_{i_n}| \big], \tag{6}$$

where $i_1, i_2 \cdots, i_n$ refer to the states with energy intersects at the inflexion [4].[1] The matrix has the form of (4), thus the concurrence can be calculated by (5).[2]

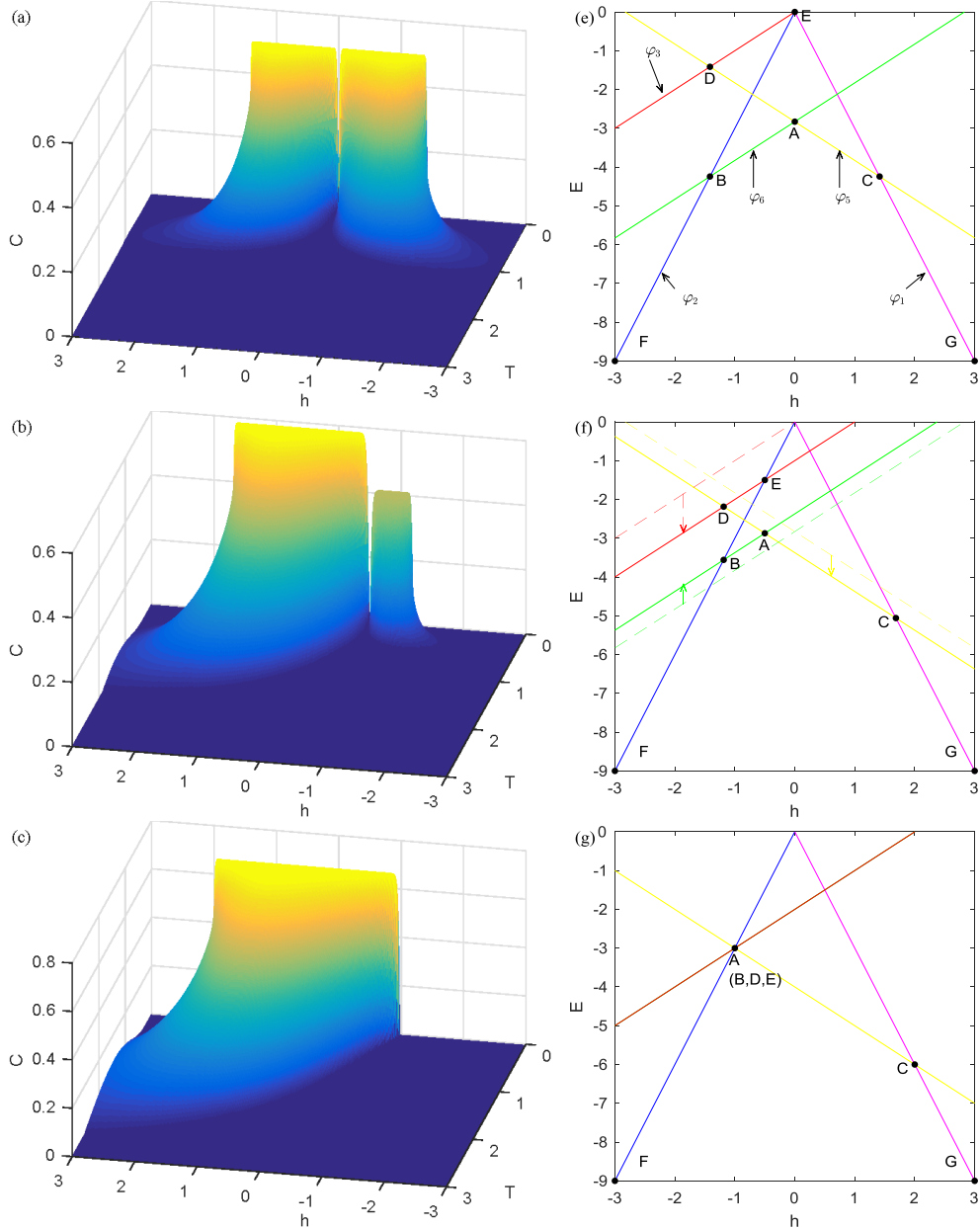

---

[1] The superscript of $C^{(mn)}$ in (2) and $\rho_{13}^{(i_1 i_2 \cdots i_n)}$ in (6) do not refer the same thing.
[2] $u, v, y, Z$ are not equal to (4a, 4b, 4c, 4d) here.

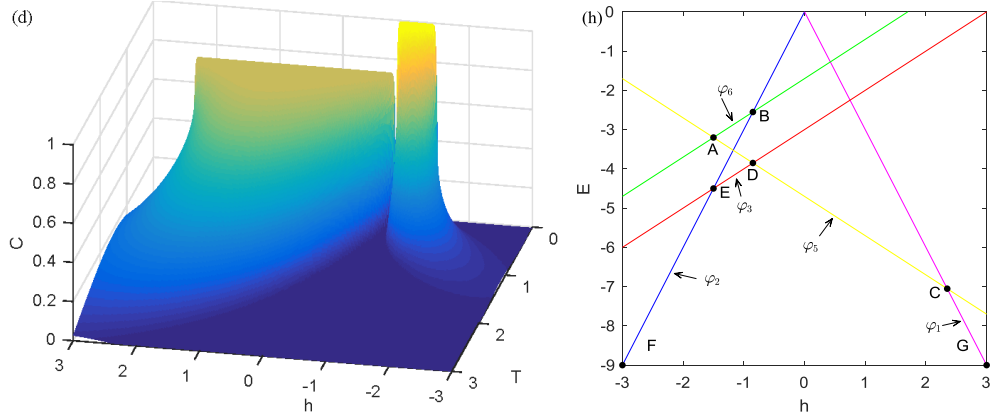

Fig. 2: (Color online) (a), (b), (c) and (d) demonstrate concurrence $C$ as function of magnetic field $h$ and temperature $T$ when $k = 0, 0.5, 1, 1.5$, respectively ($J = 1$). (e), (f), (g), (h) are the energy level diagrams correspond to the condition in the image on their left. Energy level $\varepsilon_4, \varepsilon_7, \varepsilon_8$ are omitted as they cannot serve as ground states at non-negative $k$. In (e) and (f) the ground state broken line is F-B-A-C-G, with ground state $\varphi_2, \varphi_6, \varphi_5, \varphi_1$ on each segment. Dashed lines in (f) refer to the energy levels at (e), the arrows indicate the direction energy levels translate as $k$ increases. In (h), ground state broken line changes to F-E-D-C-G, $\varphi_3$ replaces $\varphi_6$ as part of the ground state.

When $k = 0$, the ground state forms the broken line F-B-A-C-G, as shown in Fig. 2(e). Radial BF and CG correspond to state $\varphi_2$ and $\varphi_1$, respectively, whose concurrences are both 0. Segment AB represents $\varphi_6$ while AC represents $\varphi_5$. $C_6 = C_5 = 0.5$, as mentioned before. The inflexion point A is in a mixed state of $\varphi_5$ and $\varphi_6$. By calculating the concurrence using $\rho_{13}^{(56)}$ and (5), we can find that the alternate qubits are unentangled at A point. In a word, the concurrence makes two identical square-wavelike shape with infinitesimal interval on the plane $T \to 0$ due to transformation of ground state. The shape smooths out at higher temperature, generating the boot-like structures, with height $C_+ = C_5$ and $C_- = C_6$.

According to (2c), (2e) and (2f), $\varepsilon_3$ and $\varepsilon_5$ decrease while $\varepsilon_6$ increases as $k$ rises from 0 to 1. This causes line AC and DE to descend and AB to ascend, making segment AB shorten and AC extend, as shown in Fig. 2(f). Demonstrated in Fig. 1, at the same time $C_5$ goes up and $C_6$ down. Synthesizing all these factors we can explain the inflation and shrinkage of the two boots. The position of the dip can be located by computing the abscissa of point A as a function of $k$

$$h_{dip} = -k. \qquad (0 \leq k < 1) \quad (7)$$

The magnitude of the dip figured up from $\rho_{13}^{(56)}$ and (5) is

$$C_{dip} = \frac{1}{2} - \frac{\sqrt{2}}{\sqrt{k^2+8}}, \qquad (0 \leq k < 1) \quad (8)$$

which is a monotonically increasing function of non-negative $k$.

Line AB ascends while DE descends. They coincide at $k = 1$. This moment segment AB

collapses to a point, which is also the intersection of the four energy levels $\varepsilon_2$, $\varepsilon_3$, $\varepsilon_5$ and $\varepsilon_6$, leading to the disappearance of the negative boot and appearance of a point of four-fold degeneracy. The dip can be considered to have vanished, then $C_{dip}$ could be set to zero. Or it could be assigned to locate at the abscissa of the intersection point A for continuity with (7), i.e. $h_{dip} = h_A = -1$. Then the concurrence can be calculated using $\rho_{13}^{(2356)}$ and (5), which is exactly 0.

The peak with consistent height $C_- = 1$ is an outcome of the substitution of $\varphi_3$ for $\varphi_6$ as part of the ground state when $k$ exceeds 1. As shown in Fig. 2(h), the broken line now becomes F-E-D-C-G, with the dip locates at the abscissa of point D instead of A, i.e.

$$h_{dip} = h_D = \frac{1}{2}\left(k - \sqrt{k^2 + 8}\right). \qquad (k > 1) \quad (9)$$

Calculate the limitation of (9) and compare the expression of $\varepsilon_3$ and $\varepsilon_5$ in (2c) and (2e), it can be inferred that the abscissa of D approaches 0 but never passes it as $k \to \infty$, and the ground energy will form an approximately symmetric pattern about the line $h = 0$. Moreover, notice that $C_5$ approaches 1 for infinite $k$, the two boots return to a seemingly symmetric arrangement instead of one suppresses the other. Fig. 3(a) plots the concurrence as function of $T, h$ at $k = 10$, which implies this tendency to some extent. The state of the alter qubits at the dip turns to $\rho_{13}^{(35)}$, with concurrence

$$C_{dip} = \frac{1}{4}\left(1 - \frac{k}{\sqrt{k^2+8}}\right). \qquad (k > 1) \quad (10)$$

This is a monotonically decreasing function when $k > 1$.

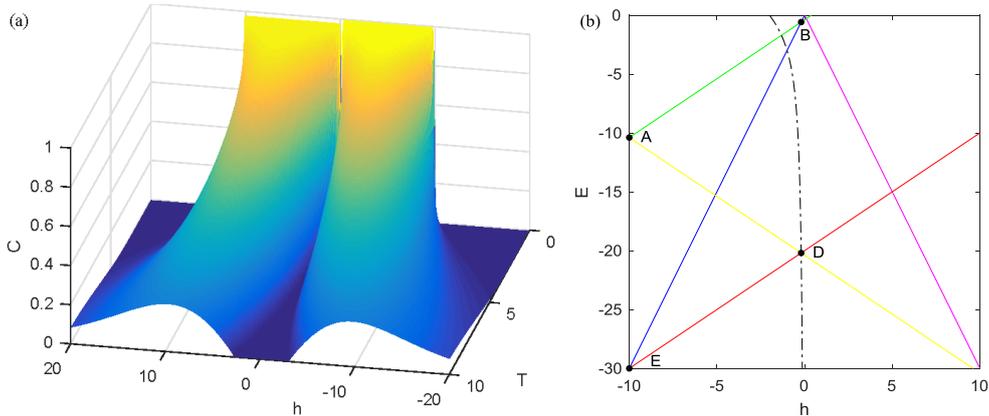

Fig. 3: (Color online)  (a) Concurrence $C$ as function of $T, h$ at $k = 10$. (b) Corresponding energy level diagram of (a). Dash-dotted line is the orbit of point D as $k$ varies. It has the asymptote $h = 0$.

Utilizing the characteristics of the ground state and its related expressions obtained above, we can analyze the behavior of the concurrence when temperature $T$ approaches 0 instead of calculating the $T \to 0$ limit of (5) directly, which is complicated and usually has no analytic solution.

The state of the positive boot at $T \to 0$ remains $\varphi_5$. The curve of the height $C_+(k) = C_5$ is shown in Fig. 1. The state of the negative boot at $T \to 0$ changes from $\varphi_6$ to none then to $\varphi_3$ as $k$ increases, making height $C_-$ a discontinuous piecewise function composed of $C_6$, 0 and 1, as plotted in Fig. 4(a).

Now combining (8), (10) and $C_{dip}\big|_{k=1} = 0$, we can get a function of the magnitude of the dip $C_{dip}$ with variable $k$, which is plotted in Fig. 4(b). Mutation occurs at $k = 1$ from $\frac{1}{2} - \frac{\sqrt{2}}{3}$ to 0 then $\frac{1}{6}$. This is caused by the change of the energy levels that cross, as mentioned above.

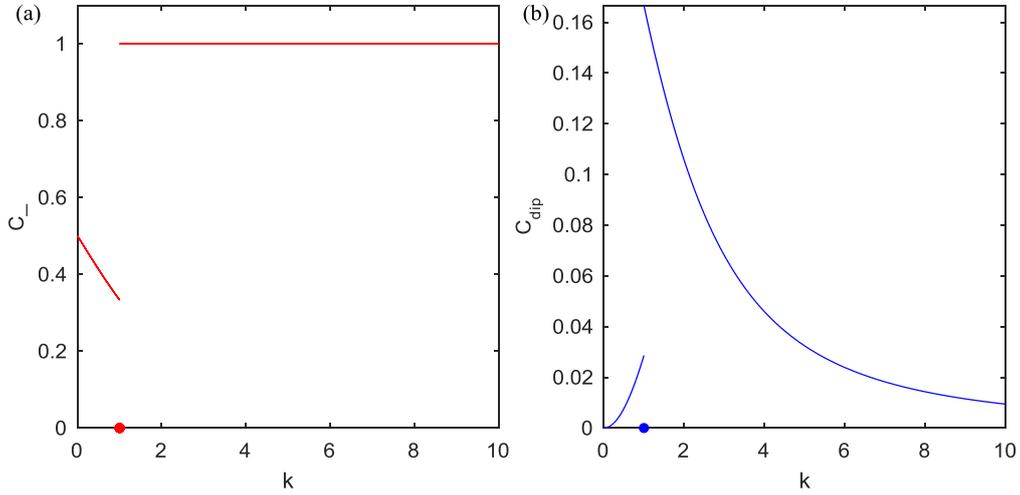

Fig. 4: (Color online) (a) Negative boot's height $C_-$ as function of $k$. (b) Value of the dip $C_{dip}$ as function of $k$.

We can find that the concurrence experiences a sudden change when the three-spin interaction $k$ passes from left limit $1_-$ to right limit $1_+$. Besides, note that the negative boot degrades from square wavelike to a δ-function when $k = 1_-, 1_+$. It adjoins the positive boot to form the dip, which also behaves as a δ-function, so the change just appears around $h_{dip}\big|_{k=1} = -1$. Fig. 5(a), (b) and (c) exhibit the mutation. Remember that the two-spin interaction $J$ has been set as the unit, $k = 1$ represents $k = J$ actually. In conclusion, the sudden change appears when one of the value of $k$ and $J$ passes the other when the magnetic field is around $h = -k$.

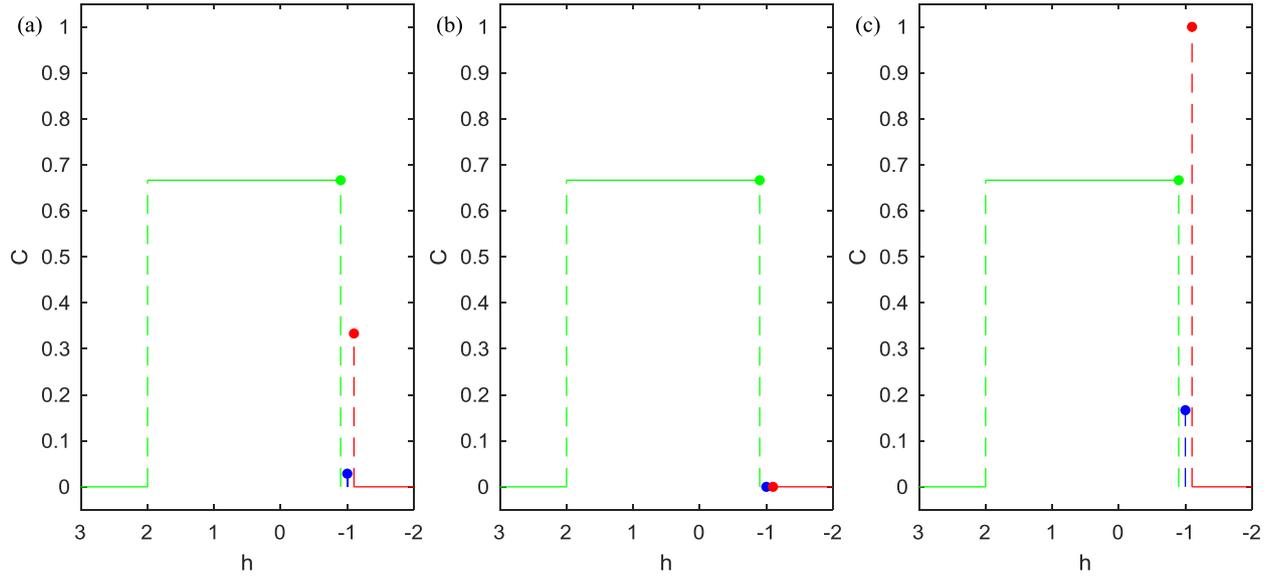

Fig. 5: (Color online) (a), (b) and (c) refer to concurrence $C$ as function of $h$ when $k = 1_-, 1, 1_+$, respectively. In each graph, from left to right the three points locate at $h = 1_-, 1, 1_+$. The right most point represents the $\delta$-function collapsing from the negative square wavelike function.

Inspecting the energy and concurrence in (2), we find symmetry with respect to $k$. Calculation confirms that the concurrence behaves mirrorlike when $k < 0$ comparing to its opposite number, as plotted in

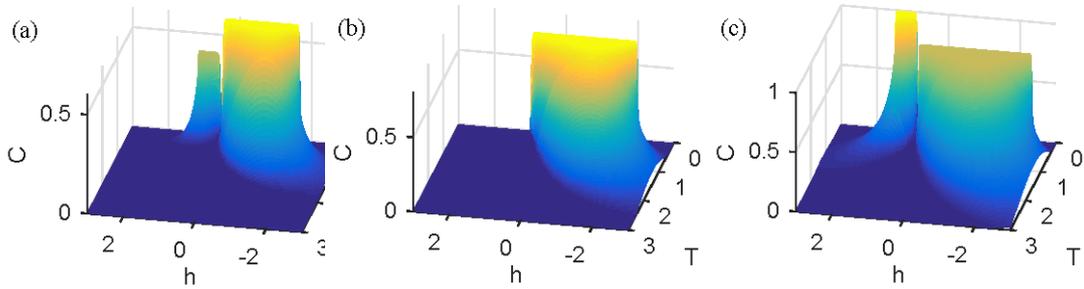

Fig. 6. When the absolute value of $k$ equals, the concurrence is symmetrical about $h = 0$. The sudden change happens when $k = -J$ near the magnetic field $h = -k$. $h$ shares the same equation with $k \geq 0$ condition.

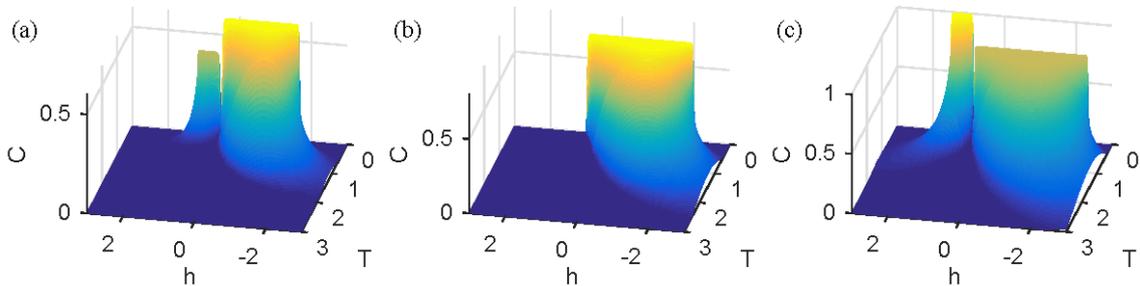

Fig. 6: (Color online) (a), (b) and (c) are plotted at k = $-0.5, -1, -1.5$, respectively. Note they are the mirror of Fig. 2 (b), (c) and (d) about h = 0. The mutation happened around k = $-1$, h = 1.

### IV. Effect of $k$ on nearest-neighbor qubits

The concurrence of the nearest-neighbor qubits can be analyzed with identical procedure mentioned in the paper. The main points are shown in

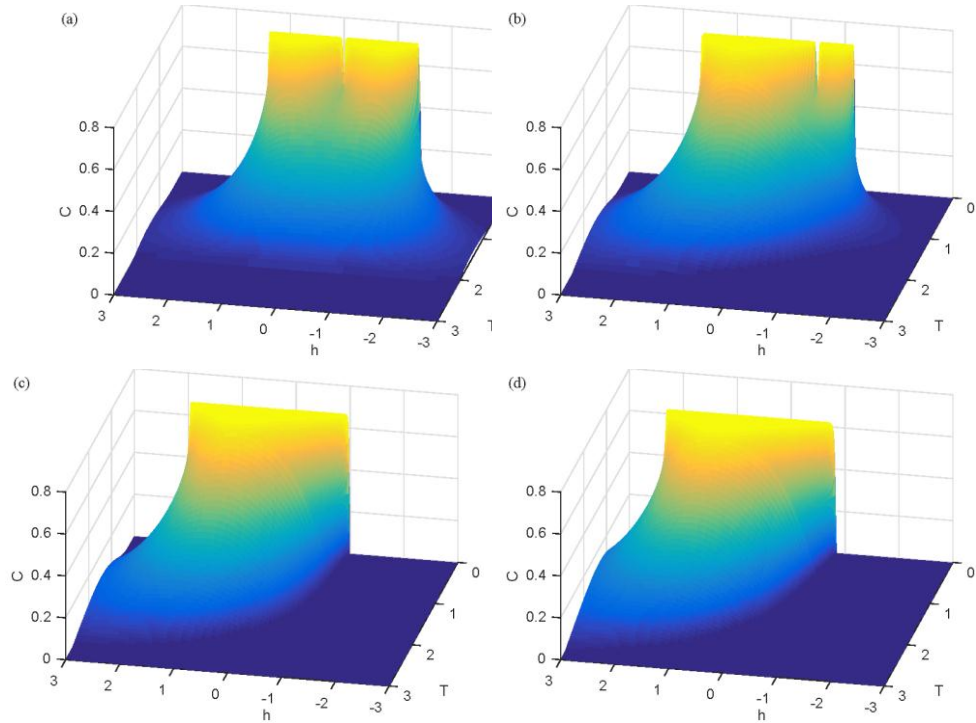

Fig. 7 and Fig. 8. When $k \geq 0$, conclusions include: qubit pair 12 and qubit pair 23 behave the same; the two boots joint at high temperature; negative boots never exists as long as $k \geq J$. It as well shows mirror-symmetry about the positive and negative $k$. Finally, we can find that increasing of $|k|$ suppresses the concurrence between neighbor qubits.

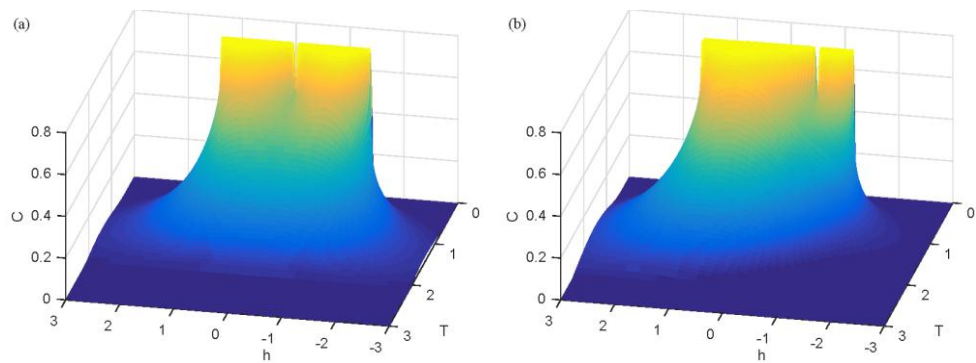

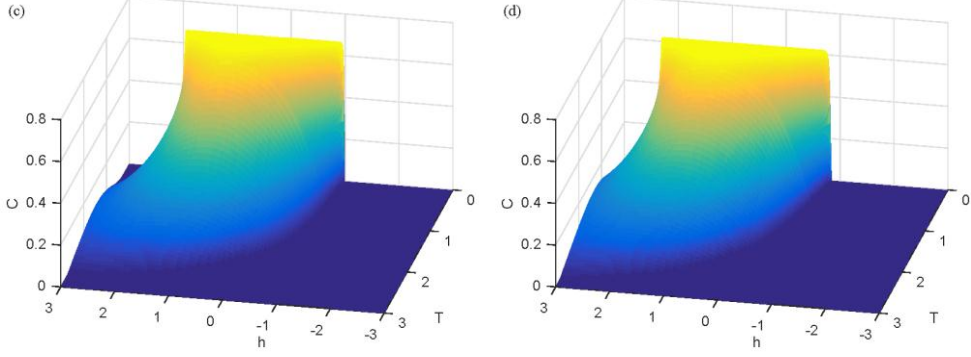

Fig. 7: (Color online) (a), (b), (c) and (d) correspond to $C^{(12)}$ at $k = 0, 0.5, 1, 1.5$, respectively. The height $C_+$ and $C_-$ keep the same and descend slowly when $0 \leq k < 1$. $k > 1$, there is no negative boot, this can be explained by $C_3^{(12)} = 0$. Thus the mutation at $k = 1$ is only vanishing of negative boot and the dip, which is analogous to Fig. 5 (a), (b).

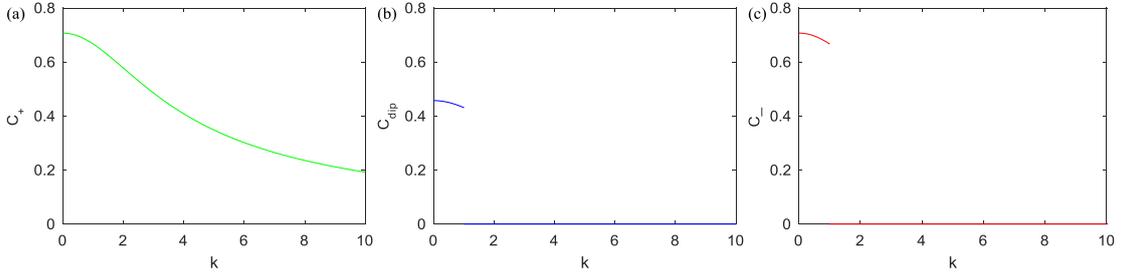

Fig. 8: (Color online) (a), (b) and (c) are $C_+, C_{dip}$ and $C_-$ as function of $k$ repectively. They all approach 0 as $k \to \infty$, which means three-spin interaction destructs the concurrence between nearest-neighbor qubits.

## V. Conclusions

In this paper, we discuss the effect of three-spin interaction on the thermal entanglement of Heisenberg XXZ model, mainly study the pairwise concurrence of the alternate qubits 13. We can find that three-spin interaction introduces a symmetry broken about the effect of magnetic field $h$. Generally speaking, it enhances the concurrence between alternate qubits while weakens the entanglement of nearest-neighbor qubits, except for the regions where the shrinking boot structure and the dip locate. The concurrences at these two regions behave rather peculiar, and a mutation occurs when one of the value of three-spin interaction $|k|$ and nearest-neighbor-spin coupling strength $|J|$ passes the other when magnetic field $h = -k$. This mutation might hold some physical meanings that deserves further researches.

**Acknowledgments**

This work is supported by the National Natural Science Foundation of China (Grant No. 11574022) and the Fundamental Research Funds for the Central Universities of Beihang

University (Grant No. YWF-17-BJ-Y-70).